\documentclass[
floatfix,
reprint,
amsmath,
amssymb,
aps,
pra,
]{revtex4-2}

\usepackage{graphicx}
\usepackage{dcolumn}
\usepackage{bm}
	
\newcommand{\ket}[1]{\left| #1 \right\rangle}
\newcommand{\bra}[1]{\left\langle #1 \right|}
\newcommand{\abs}[1]{\left|#1\right|}
\newcommand{\vb}[1]{\boldsymbol{#1}} 
\newcommand{\vu}[1]{\boldsymbol{\hat{#1}}} 

\begin{document}

\preprint{APS/123-QED}

\title{Quantum-logic spectroscopy of forbidden vibrational transitions\\ in single nitrogen molecular ions}

\author{Aleksandr~Shlykov}
 \homepage{These authors contributed equally to this work}
\author{Meissa~L.~Diouf}
 \homepage{These authors contributed equally to this work}
\author{Richard~Karl}
\author{Mikolaj~Roguski}
\author{Umesh~C.~Joshi}
\altaffiliation{}
\author{Stefan Willitsch}%
 \email{stefan.willitsch@unibas.ch}
\affiliation{%
 Department of Chemistry, University of Basel, Klingelbergstrasse 80, 4056 Basel, Switzerland.
}%

\date{\today}

\begin{abstract}
\noindent
Electric-dipole forbidden spectroscopic transitions in atoms form the basis of many advanced implementations of quantum computers, atomic clocks and quantum sensors. Coherently addressing such transitions in molecules which are among the most ubiquitous and versatile quantum objects has remained a long-standing challenge owing to their complex energy-level structure. Here, we report the search, observation and coherent manipulation of electric-quadrupole rotational-vibrational transitions in single trapped molecules using a quantum-logic-spectroscopy protocol. We identified individual hyperfine-Zeeman-rotational components of the fundamental vibrational transition of the nitrogen molecular ion, N$_2^+$, and performed coherent population transfer between energy levels. Our work opens up new perspectives for precision molecular spectroscopy, for high-fidelity qubits encoded in the rotational-vibrational motion of molecules, for precise infrared molecular clocks and for searches for new physics.
\end{abstract}

\maketitle

\noindent Electric-dipole-forbidden transitions are spectral excitations in atoms and molecules which are forbidden to first order but become allowed by higher-order terms in the interaction between matter and radiation, e.g., by electric-quadrupole or magnetic-dipole couplings. Such transitions can exhibit extremely narrow linewidths which render them the basis of a range of key technologies in the realms of quantum science and frequency metrology. In atomic systems, they form qubits in leading platforms for quantum computation~\cite{haeffner08a,bruzewicz19a,ringbauer22a}, they are at the core of today's most precise clocks~\cite{yang25a, hausser25a, marshall25a} and they find applications in advanced quantum sensors~\cite{kotler11a, baumgart16a, franke23a}.  

Electric-quadrupole rotational-vibrational (rovibrational) transitions in molecules have been predicted to possess exceedingly high quality factors with linewidths down to the nHz level in the infrared spectral domain \cite{kajita14a, germann14a, hanneke16a, korobov18a, najafian20b, wolf24a, ubachs25a}. Due to the properties of the molecular energy-level structure, such transitions can offer excellent systematics and resilience against external perturbations~\cite{karr08a, zelevinsky08a, kajita15a, najafian20b, hanneke20a, leung23a, wolf24a}. In this context, forbidden vibrational transitions in homonuclear diatomic molecular ions such as H$_2^+$, N$_2^+$, O$_2^+$ and I$_2^+$ have been proposed as prime candidates for the implementation of highly precise mid-infrared frequency standards, for the realization of ultrahigh-fidelity qubits, for accurate determinations of the values of fundamental physical constants and for probing physics beyond the standard model~\cite{schiller05a, flambaum07a, karr14a, schiller14a, kajita14a, najafian20b, safronova18a, hanneke20a, magde24a}. 

However, precision spectroscopy of forbidden vibrational transitions is challenging on several levels~\cite{germann14a, schenkel24a}. Their narrow linewidth invariably implies extremely weak line strengths (about ten orders of magnitude weaker than typical dipole-allowed infrared transitions~\cite{korobov18a, najafian20b}) which require highly sensitive experimental methods. Moreover, determining the positions and strength of extremely weak spectral lines poses a demanding search problem~\cite{furst20a, chen24a, cheung25a, yu26a}, particularly in multi-electron molecules for which the line positions cannot be predicted theoretically with sufficient accuracy \cite{najafian20b, wolf24a}. Electric-dipole forbidden rovibrational transitions in homonuclear ions have recently been observed for the first time in N$_2^+$ \cite{germann14a} and subsequently in H$_2^+$ \cite{schenkel24a, alighanbari25a} using chemical detection techniques such as laser-induced charge transfer and resonance-enhanced multiphoton dissociation. These chemical methods destroy the molecules in every experimental cycle, necessitating reloading and reinitialization of the sample. This severely limits the experimental duty cycle and, therefore, the sensitivity and precision of the measurements \cite{germann14a, alighanbari25a}. 

In recent years, coherent methods have been developed for the non-destructive manipulation and readout of quantum states of single molecules~\cite{wolf16a, chou17a, sinhal20a, holzapfel25a}. These protocols can be seen as variants of the `quantum-logic spectroscopy' (QLS) originally conceived in the development of the Al$^+$ ion optical clock~\cite{schmidt05a}. The general approach relies on projecting information about the quantum state of a `spectroscopy' ion onto a motional mode shared with a `logic' ion in a trap, from which it can be read out sensitively using resonance fluorescence. These inherently non-destructive techniques enable quantum-non-demolition (QND) measurements which leave not only the chemical nature, but also the quantum state of the molecule intact~\cite{wolf16a, sinhal20a, liu24a}. Thus, the sensitivity of the experiments can be improved by several orders of magnitude compared to traditional chemical readout techniques, although only a single particle is probed at a time instead of large ensembles~\cite{sinhal20a, chou20a}.

\begin{figure*}
    \centering
    \includegraphics[width=1.0\textwidth]{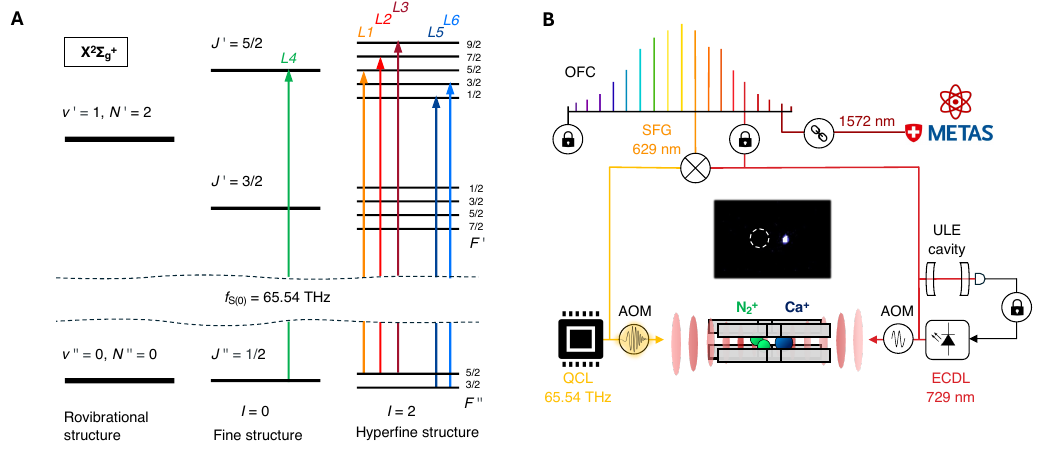}
    \caption{
    \textbf{Details of the experiment.} 
    \textbf{(A)} Energy levels and spectroscopic transitions of N$_2^+$ relevant for the current study.  Coloured arrows indicate (hyper)fine components of the S(0) rovibrational band observed in this work. $v, N, J, I$ and $F$ stand for the vibrational, rotational, spin-rotational, nuclear-spin and total-angular-momentum quantum numbers.  
    \textbf{(B)} Schematic of the experiment, see text for details. Abbreviations: QCL - quantum cascade laser; OFC - optical frequency comb; ECDL - external-cavity diode laser; ULE - ultralow expansion; SFG - sum-frequency generation; AOM - acousto-optic modulator; METAS - Swiss Federal Institute of Metrology.
    }
    \label{fig:schemes}
\end{figure*}

Here, we report the application of QLS methods to the study and coherent manipulation of forbidden rovibrational transitions in single molecular ions, opening up a new frontier in the realm of molecular spectroscopy. We applied a QLS protocol~\cite{sinhal20a} in conjunction with coherent population transfer by rapid adiabatic passage (RAP)~\cite{wunderlich07a, noel12a, chen24a} to the search, identification and spectroscopy of electric-quadrupole transitions in the nitrogen molecular ion N$_2^+$. The high sensitivity of our approach enabled the recording of overview spectra of the hyperfine structure and the identification of individual (hyper)fine components of rovibrational transitions in single trapped N$_2^+$ molecules with high accuracy, improving the previously reported fundamental vibrational frequency by an order of magnitude.

Specifically, we studied individual hyperfine-Zeeman components of the $\ket{v''=0, N''=0}\rightarrow\ket{v'=1, N'=2}$ transition, i.e., of the S(0) rotational line of the fundamental infrared transition in the $X~^2\Sigma_g^+$ electronic ground state of N$_2^+$ originating from the rotational ground state (Fig. \ref{fig:schemes}A). This level consists of 12 different Zeeman states associated with the $J''=1/2$ and $F''=1/2,3/2$ fine and hyperfine components of the $I=0$ and $I=2$ nuclear-spin isomers, respectively. Here, $v,N, J$ and $F$ denote the vibrational, rotational, fine and hyperfine quantum numbers and the (double) prime stands for the (lower) upper states of spectroscopic transitions. 

A simplified schematic of the experiment is shown in Fig.~\ref{fig:schemes}B, see~\cite{methods} and~\cite{karl26a} for further details. Single N$_2^+$ ions were trapped and cooled together with single Ca$^+$ ions in a linear radiofrequency ion trap to form two-ion Coulomb crystals (image inset in Fig. \ref{fig:schemes}B). A quantum-cascade laser (QCL) operating at a wavelength around 4.57~$\mu$m (65.54 THz) was employed to excite the target spectroscopic transitions. The QCL was locked to an optical frequency comb (OFC) which was disciplined to a high-finesse cavity resulting in linewidths at the sub-kilohertz level~\cite{sinhal23a}. The OFC was referenced to the Swiss primary frequency standard disseminated by the Swiss National Institute of Metrology (METAS) via a stabilized optical-fiber link enabling SI-traceability of measured frequencies~\cite{husmann21a}. Spectroscopic transitions were driven by a RAP protocol using frequency-chirped laser pulses~\cite{malinovsky01a} from the QCL, produced by an acousto-optical modulator (AOM). This approach enabled efficient, fully reversible population transfer between the levels involved in the transitions without requiring precise knowledge of the resonance frequencies. Simultaneously, it allowed the recycling of the same single molecule in sequential spectroscopic experiments which was crucial for achieving high experimental duty cycles and measurement statistics. In this way, the positions of unknown spectral lines could be detected with an uncertainty limited only by the width of the RAP frequency sweep of the laser pulse. 

\subsection*{Quantum-logic spectroscopy protocol}
\label{sec:QLS_protocol}

\begin{figure*}
  \includegraphics[width=\linewidth]{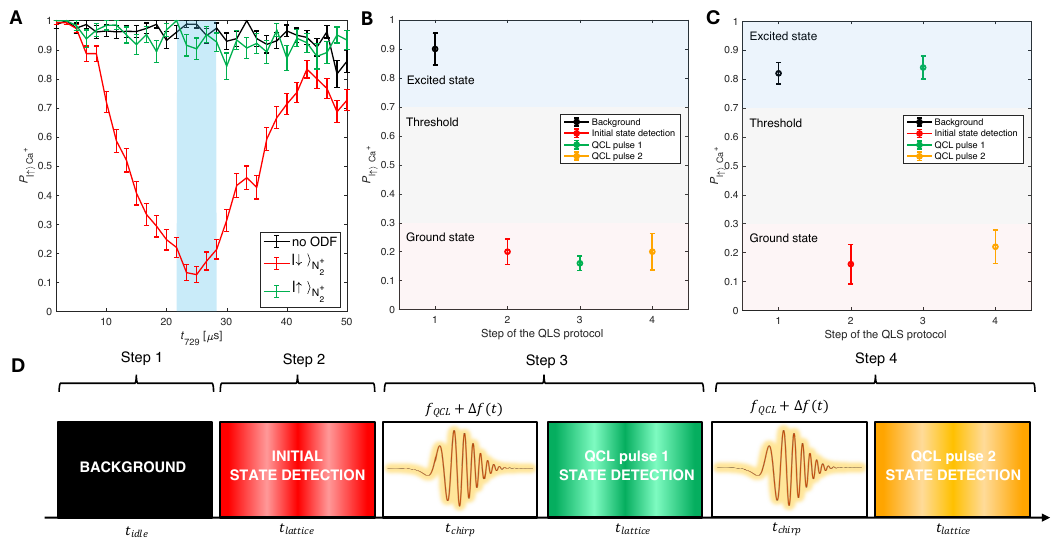}
  \caption{
  \textbf{Quantum-logic spectroscopy protocol.} 
  \textbf{(A)} Population $P_{\ket{\uparrow}_{\text{Ca}^+}}$ of the excited state of Ca$^+$ as a function of 729~nm laser-pulse duration $t_{729}$ during Rabi flops on a blue motional sideband of the clock transition $\ket{\uparrow}_{\text{Ca}^+}\rightarrow\ket{\downarrow}_{\text{Ca}^+}$ in Ca$^+$ following coherent motional excitation of a Ca$^+$-N$_2^+$ two-ion string by a state-dependent optical dipole force on N$_2^+$. The red and green traces show experiments with N$_2^+$ in its rovibrational ground ($\ket{\downarrow}_{\text{N}_2^+}$) and an excited state ($\ket{\uparrow}_{\text{N}_2^+}$), respectively. The black trace represents the signal obtained without the application of an ODF. The blue-shaded area indicates the interval of 729~nm laser pulse lengths for which maximum detection contrast was achieved. Error bars represent the standard error of the mean of the binomial distribution of 150, 75 and 150 measurements for the red, green and black traces. 
  \textbf{(B)} `Negative' and \textbf{(C)} `positive' outcomes of the quantum-logic protocol for the excitation of a dipole-forbidden infrared transition. State-detection results after every step of the sequence are indicated by the colored dots, and represent an average of 50 state-detection measurements. The colors of the dots match the steps in panel \textbf{(D)} which illustrates the quantum-logic protocol used for the spectroscopy of N$_2^+$ ion. See text for details.}
  \label{fig:QLS_sequence}
\end{figure*}

\noindent A recently developed method for QND state detection of molecular ions was implemented for quantum logic spectroscopy (see \cite{sinhal20a} and \cite{methods} for details). Briefly, the N$_2^+$ - Ca$^+$ two-ion Coulomb crystals were laser-cooled to the ground state of their common center-of-mass motional mode by resolved-sideband techniques. A state-dependent optical dipole force (ODF), generated by a 1D traveling optical lattice consisting of two counter-propagating laser beams coherently excited the motion of the atomic-molecular two-ion crystal in the trap if and only if the N$_2^+$ ion was in the rovibrational ground state $\ket{\downarrow}_{\mathrm{N}_2^+}$. The resulting motional excitation of the ion string was probed using Rabi spectroscopy on a blue motional sideband of the $\ket{\downarrow}_{\text{Ca}^+}$~[4s~$^2$S$_{1/2}(m_j=-1/2)]\rightarrow \ket{\uparrow}_{\text{Ca}^+}$~[3d~$^2$D$_{5/2}(m_j=-5/2)$]  clock transition in Ca$^+$ at 729~nm. The red trace in Fig.~\ref{fig:QLS_sequence}A shows a typical Rabi flop obtained if the N$_2^+$ ion was in its rovibrational ground state $\ket{\downarrow}_{\mathrm{N}_2^+}$. By contrast, the green trace represents the signal obtained for an ion in an excited rovibrational state $\ket{\uparrow}_{\mathrm{N}_2^+}$, which is effectively indistinguishable from the background signal level acquired without application of the optical lattice (black trace). For the purpose of reliable state detection, it was not necessary to record the entire Rabi flop, but only the signal level at the Rabi time of maximum contrast indicated by the blue-shaded area in Fig.~\ref{fig:QLS_sequence}A. The fidelity for detecting the molecular ion in the rovibrational ground state was demonstrated to be $>99$\%~\cite{sinhal20a, roguski25a}.

The experimental protocol used for quantum-logic spectroscopy and coherent control of a single N$_2^+$ ion is illustrated in Fig.~\ref{fig:QLS_sequence}D. Following state preparation, each experimental cycle started with a background measurement without application of the optical lattice (Step 1, black trace in Fig.~\ref{fig:QLS_sequence}A). This step served to establish the background signal level for discrimination of the rovibrational ground state of the molecular ion from excited states. 
Subsequently, an initial QND state detection (Step 2) was carried out to verify that the molecule was prepared in the $\ket{\downarrow}_{\mathrm{N}_2^+}$ state. A first RAP experiment was then performed using a QCL pulse with a linear frequency sweep across an interval $\Delta f$ and a duration $t_{chirp}$ to excite the molecule to $\ket{\uparrow}_{\mathrm{N}_2^+}$ if a spectroscopic transition from $\ket{\downarrow}_{\mathrm{N}_2^+}$ was present within the relevant interval (Step 3). This was followed by a QND state-detection measurement to verify the transfer of population to $\ket{\uparrow}_{\mathrm{N}_2^+}$ in the case of a successful excitation. Eventually, an identical chirp and state-detection sequence (Step 4) was employed to transfer the population back to  $\ket{\downarrow}_{\mathrm{N}_2^+}$ and re-initialize the molecule.

Typical outcomes of this protocol are shown in Fig.~\ref{fig:QLS_sequence}B,C, where each data point corresponds to the result of one of the four steps discussed above. The shaded areas indicate the signal level when the molecular ion was identified as being in $\ket{\downarrow}_{\mathrm{N}_2^+}$ (red-shaded area) or in a rovibrationally excited state  $\ket{\uparrow}_{\mathrm{N}_2^+}$ (blue-shaded area). These levels were calibrated by a series of previous QND state-detection experiments~\cite{sinhal20a}. 
If N$_2^+$ did not change its state because no spectroscopic transition was present within the relevant frequency-sweep interval (Fig.~\ref{fig:QLS_sequence}B), the outcome was designated as `negative'. If N$_2^+$ changed its state twice in consecutive RAP experiments, being excited by a first pulse and then returning to the ground state in subsequent pulses (Fig.~\ref{fig:QLS_sequence}C), it was classified as `positive' and counted as a successful identification of a spectroscopic transition within the frequency interval considered.

\subsection*{Characterization of coherent transfer of rovibrational population with RAP}
\label{sec:RAP}

\begin{figure}
    \centering
    \includegraphics[width=0.5\textwidth]{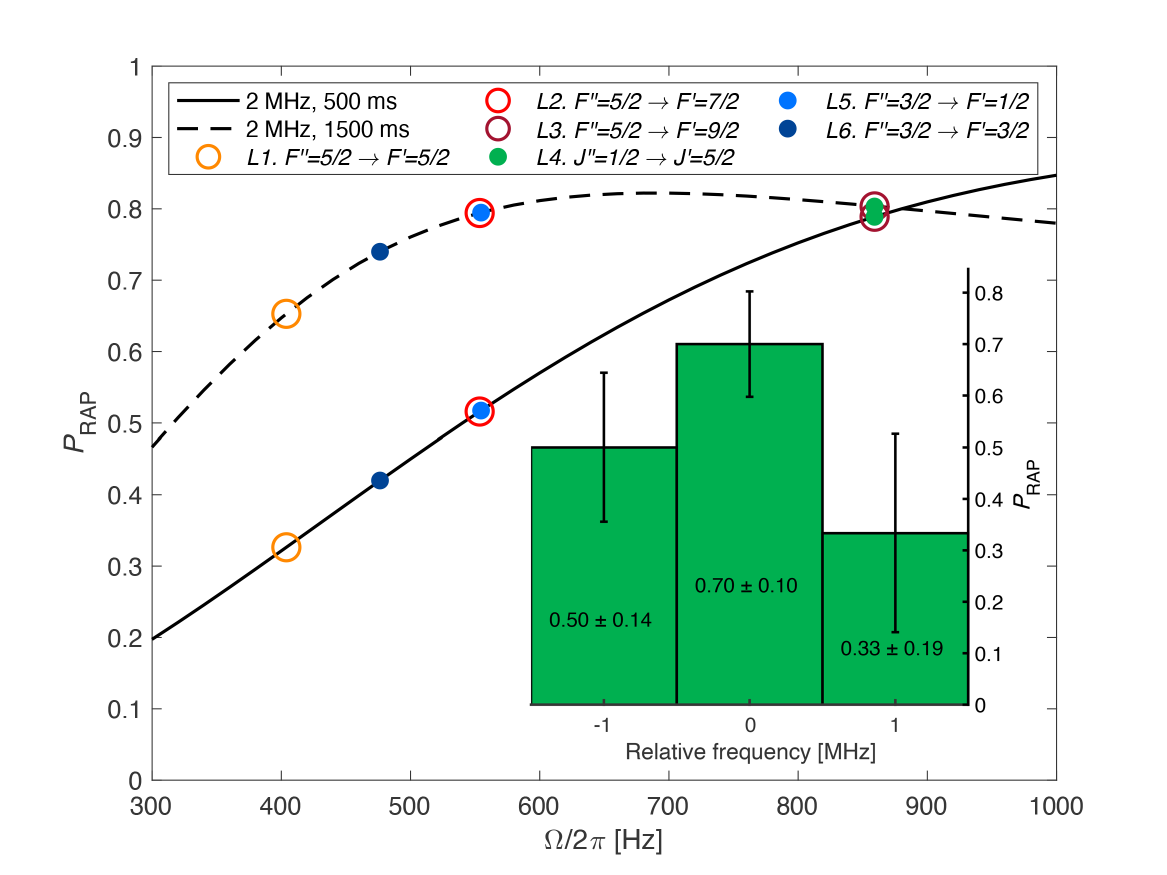}
    \caption{\textbf{Probability of population transfer.} Probability for coherent population transfer using rapid adiabatic passage as a function of the Rabi frequencies $\Omega$ of the transition for two different sets of RAP parameters (chirp bandwidth and duration) used in the experiment (dashed and solid lines). Coloured dots and circles represent the calculated probability of population transfer for the strongest Zeeman components of the relevant (hyper)fine transitions (color coded as in Fig.~\ref{fig:schemes}).
    \textbf{Inset:} Results of an experiment on a single N$_2^+$ ion indicating the experimental probability of population transfer on the $\ket{J''=1/2, m_F=\pm1/2} \rightarrow \ket{J'=5/2,m_F=\pm5/2}~(L4)$ transition (green dots in main figure). Values inside the bins denote experimental state-changing probabilities in the relevant frequency interval. Error bars indicate the standard errors of a binomial distribution.} 
    \label{fig:RAP}
\end{figure}

\noindent Fig.~\ref{fig:RAP} shows the probability of population transfer using RAP, calculated by solving the system of optical Bloch equations~\cite{foot05a, bloch46a} as a function of the Rabi frequency $\Omega$ of the relevant electric-quadrupole rovibrational transitions for two different sets of RAP parameters used in the experiment. The RAP parameters $\Delta f=2$~MHz, $t_{chirp}=500$~ms (solid black line) were used to address the strongest Zeeman components of the $\ket{J''=1/2} \rightarrow \ket{J'=5/2}$~(indicated as $L4$ in Fig.~\ref{fig:schemes}A) and $\ket{F''=5/2} \rightarrow \ket{F'=9/2}~(L3)$ transitions. The second set of parameters with a slower chirp rate $\Delta f = 2$~MHz, $t_{chirp}=1500$~ms (dashed black line) was implemented to drive the transitions $L1, L2, L5, L6$ (Fig.~\ref{fig:schemes}A) with comparable efficiency. The population-transfer probabilities for the strongest Zeeman components of the (hyper)fine transitions at the laser intensity employed in the present experiments are indicated with colored circles and dots in Fig.~\ref{fig:RAP} (see \cite{methods} for details of the calculation).

The inset in Fig.~\ref{fig:RAP} displays the result of a representative experiment of population transfer by RAP with the first set of parameters on the strongest Zeeman components of $L4$ $(m_F\pm1/2\rightarrow m_F\pm5/2)$ which are not resolved in the current RAP frequency-sweep interval. In this particular experiment, the N$_2^+$ ion changed state 22 times after interacting with a total of 38 chirped IR pulses. The probability of a rovibrational population transfer for the frequency interval with the highest number of successful transitions (set at a relative frequency of 0 MHz) is $70\pm10$~\%, and the total probability across all three intervals shown is $58\pm8$~\%. 
The observed population-transfer probability of $70\pm10$~\% for the central interval is compatible with the theoretical probability of 78.9~\% calculated for this transition (green dot on the solid black line).

\subsection*{Quantum-logic spectroscopy of dipole-forbidden transitions}
\label{sec:spectroscopy}

\begin{figure*}
    \centering
    \includegraphics[width=1.0\textwidth]{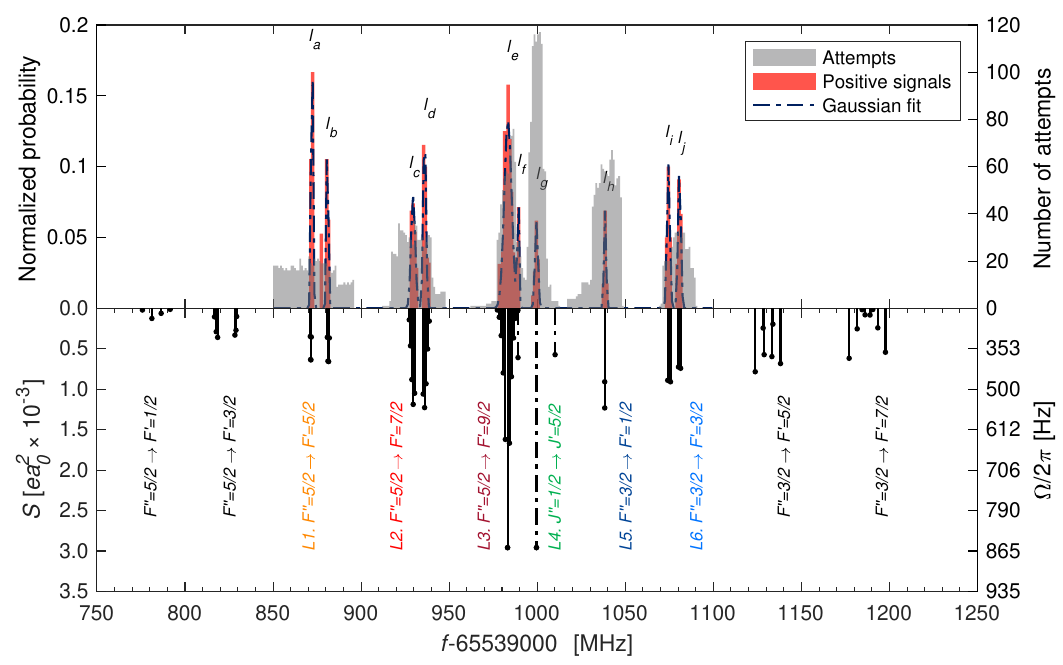}
    \caption{\textbf{Experimental spectrum and simulation.}   
    \textbf{(Top)}~Rapid-adiabatic-passage quantum-logic spectrum of the S(0) rovibrational band of N$_2^+$. The red histogram bars indicate successful population transfer by RAP normalized to the number of attempts (grey bars) within each frequency bin.   
    The dashed lines represent Gaussian fits to the data. 
    \textbf{(Bottom)}~Simulated line strengths $S$ and Rabi frequencies $\Omega$ of individual Zeeman-(hyper)fine components of the S(0) rovibrational line (labels are color-coded as in Fig.~\ref{fig:schemes}). See text for details.}
    \label{fig:main_spectrum}
\end{figure*}

\noindent The spectrum resulting from a RAP-QLS survey scan in the frequency region of the S(0) rotational component of the infrared fundamental of N$_2^+$ is displayed in Fig.~\ref{fig:main_spectrum}. The top panel shows a histogram of normalized population transfer probabilities (red bars) $-$ the ratio of `positive' RAP experiments per total number of attempts within a specific frequency bin (gray bars). The intervals feature a step size of 1 MHz and a width of 2 MHz, corresponding to the frequency-sweep width used in the RAP experiments, which was also adopted as a conservative estimate of the uncertainty of the spectral resolution of the experiment. The data were fitted using a Gaussian model \( ae^{-(f-f_0)^2/2\sigma^2} \) (blue-dashed line) to extract the center frequency of the observed spectral lines as reported in table~\ref{tab:extracted_lines}.

Fig.~\ref{fig:main_spectrum} is a summary of experiments on several hundred single molecules whose populations were initially distributed across the 12 hyperfine-Zeeman levels of the rovibrational ground state. The spectrum shows different (hyper)fine transitions of the S(0) rotational component of the fundamental vibrational band with a partially resolved Zeeman structure at an applied magnetic field of 4.7~G. The lines are labeled as transitions in the angular-momentum quantum numbers $J$ and $F$ and color-coded according to the scheme of Fig.~\ref{fig:schemes}A.

The inverted stick spectrum in Fig.~\ref{fig:main_spectrum} represents a simulation of the positions and line strengths $S$ (as defined in \cite{najafian20b} and \cite{methods}) of the transitions based on an effective Hamiltonian including hyperfine and Zeeman structure (see~\cite{methods} for details about the simulation and spectroscopic constants used). Only quadrupole transitions with the magnetic selection rule $\Delta m_F=\pm 2$ were taken into account in the simulation, where $m_F$ represents the magnetic quantum number associated with the total angular momentum $F$. This selection rule is imposed by the linear polarization of the QCL radiation and the beam propagation direction which are both perpendicular to the quantization axis of the experiment.  Note that the experimentally observed line intensities in the spectrum do not exactly reflect the theoretical line strengths $S$, but also depend on RAP efficiencies, the number of transitions that fall within a specific RAP interval and propensities in the initial state preparation of the ions. A simulation of the spectrum taking into account these factors is presented in fig.~\ref{fig:rapspectrumsimulation}.
Table~\ref{tab:lines_splitting} lists the positions of the spectroscopic lines \(L1, \ldots, L6\) observed in Fig.~\ref{fig:main_spectrum} and compares the experimental line splittings with the calculated ones. The separation between the lines observed in this study is consistent within experimental uncertainties with the values calculated using previously reported (hyper)fine spectroscopic constants~\cite{scholl98a, berrahmansour91a}. This agreement enables an unambiguous assignment of all lines observed in the experimental spectrum as indicated in Fig.~\ref{fig:main_spectrum} and table~\ref{tab:extracted_lines}.

\begin{figure}
\centering
    \includegraphics[width=0.5\textwidth]{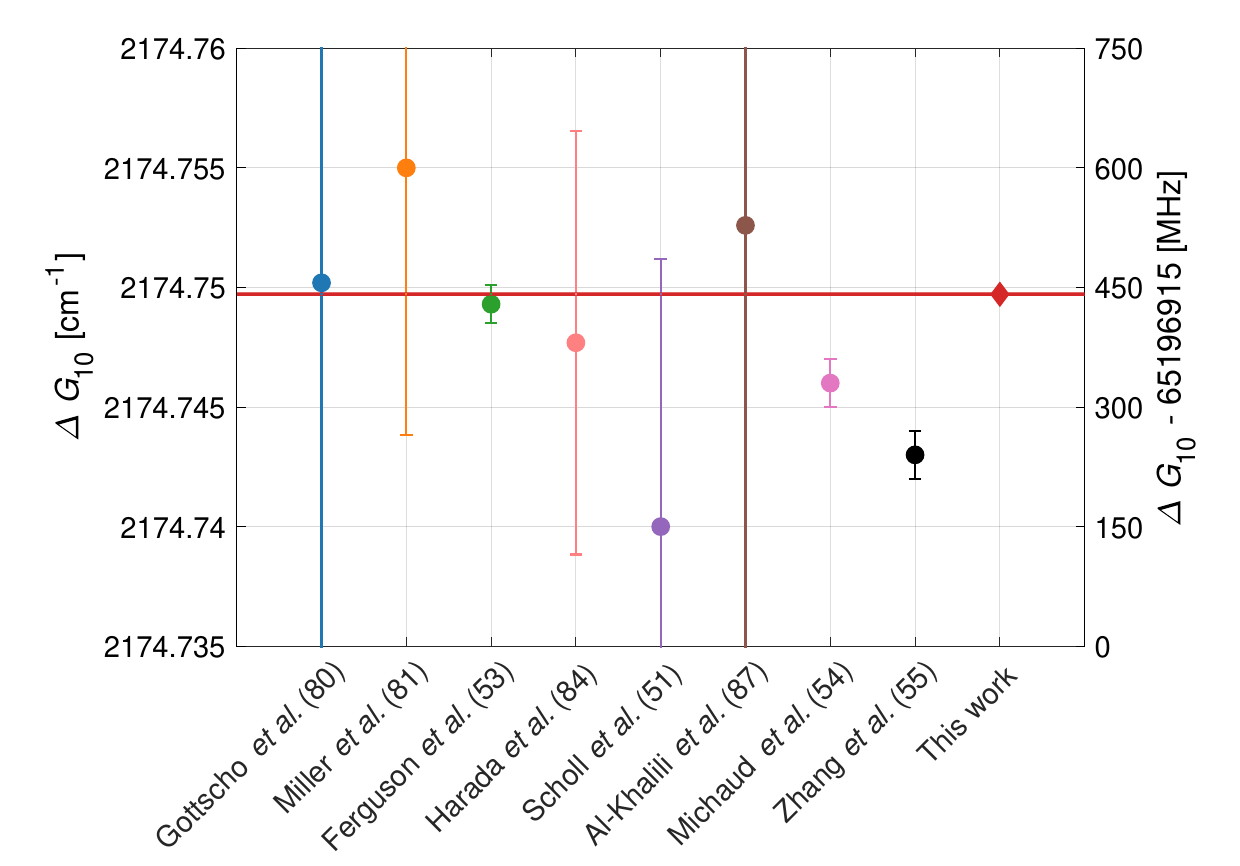}
    \caption{\textbf{Fundamental vibrational frequency $\Delta G_{10}$ of N$_2^+$.} Values reported in the literature (colored dots) compared to the present result (red diamond). For the numerical values, see tab.~\ref{tab:vib_constants}. The width of the red line represents the $1\sigma$ uncertainty of the present result, error bars indicate the uncertainties quoted in the literature.}
    \label{fig:constants}
\end{figure}

From these spectral data, an updated value of the fundamental vibrational frequency $\Delta G_{10}$ can be determined \cite{methods}. Fig.~\ref{fig:constants} shows the values of $\Delta G_{10}$ reported in the literature (colored dots) in comparison to the present result (red diamond). The most precise previous results were reported by Ferguson \textit{et al.} \cite{ferguson92a}, Michaud \textit{et al.}~\cite{michaud00a} and Zhang \textit{et al.}~\cite{zhang15a} obtained indirectly by electronic spectroscopy of the $X\rightarrow A$ and $X\rightarrow B$ electronic bands. Note that these values differ by hundreds of megahertz and within at least 2$\sigma$ uncertainty from each other. Taking into account that typical uncertainties of the rotational constant $B_1$, centrifugal distortion constant $D_1$, the effective spin-rotation constant $\gamma_1$ and the centrifugal-distortion term $\gamma_{N1}$~for the first vibrationally excited state are reported with much better precision than $\Delta G_{10}$ (see figure.~\ref{fig:mean_constants} and table~\ref{tab:mean_constants}), the fundamental vibrational frequency of N$_2^+$ can be extracted from the frequency of the $L4 $ transition in the $I=0$ isomer in Fig.~\ref{fig:main_spectrum}~as:
\begin{equation}
    \label{eq:deltagmain}
    \Delta G_{10}=f^{exp}_{L_4}-6\overline{B}_1+36\overline{D}_1-\gamma_1-6\gamma_{N1},
\end{equation}
where $\overline{B}_1, \overline{D}_1$  are weighted mean values of the corresponding spectroscopic constants reported in the literature (Supplementary Information) and $\gamma_1$ and $\gamma_{N1}$ are taken from~\cite{berrahmansour91a}, resulting in a value of $\Delta G_{10}=65197356.2(21)$~MHz or $2174.74971(7)$~cm$^{-1}$.

The present results can be compared with those reported in~\cite{germann14a} where two broad features were observed by laser-induced charge-transfer spectroscopy which were assigned to the $\ket{F''=5/2} \rightarrow \ket{F'=9/2}$,  $\ket{J''=1/2} \rightarrow \ket{J'=5/2}$ and $\ket{F''=3/2} \rightarrow \ket{F'=7/2}$ (hyper)fine components based on the spectroscopic constants reported in \cite{michaud00a}. As the present results imply a correction of the fundamental vibrational frequency reported in \cite{michaud00a} by 111.3~MHz, the original assignment of the two features observed in~\cite{germann14a} should be revisited. These lines should be attributed to the $\ket{F''=5/2} \rightarrow \ket{F'=3/2}$ and the $\ket{F''=3/2} \rightarrow \ket{F'=1/2}$ hyperfine components.

\subsection*{Prospects for precision molecular spectroscopy}
\label{sec:prospects}

\noindent In the present work, we demonstrated the application of quantum-logic techniques to the search, identification and coherent manipulation of forbidden rovibrational transitions in single nitrogen molecular ions. Using a RAP protocol for rovibrational population transfer in the mid-infrared spectral domain, we identified a range of hyperfine-Zeeman components of the S(0) rotational transition of the fundamental vibrational band of N$_2^+$. At the same time, RAP enabled the efficient state-reinitialization and recycling of the single molecules in the spectroscopy experiments, crucially increasing the duty cycle and thus the sensitivity of the experiments. The present study provided a new value of the fundamental vibrational frequency of N$_2^+$ which is an order of magnitude more precise than the most accurate previous determination based on conventional spectroscopic techniques using ensembles of molecules.

This study opens up a new frontier in molecular spectroscopy by paving the way for precision measurements in the infrared domain on single molecules. Indeed, several of the infrared quadrupole transitions observed here have been predicted to show excellent systematics and resilience against external perturbations \cite{kajita14a, kajita15a, najafian20b} which makes them prime candidates for frequency metrology. For instance, the stretched Zeeman components of the hyperfineless $L4$ transition were computed to exhibit small, linear Zeeman shifts for which measurement precisions of order $10^{-14}-10^{-15}$ can be expected considering typical magnetic field fluctuations in ion-trapping experiments. Similarly, specific hyperfine-Zeeman components of the $L3$ transition were predicted to show `magic' behavior, i.e., their resonance frequencies are independent of the magnetic field to first order at specific field strengths. These transitions offer prospects for measurement precisions on the $10^{-16}-10^{-17}$ level \cite{najafian20b}. The metrological properties of these quadrupole infrared transitions render them prime candidates for encoding high-fidelity qubits in the rovibrational motion of N$_2^+$, for the implementation of highly precise infrared molecular clocks and for probing physics beyond the standard model such as a putative time variation of the proton-to-electron mass ratio \cite{kajita14a, kajita15a, najafian20b, hanneke20a}. 


\section*{Methods}
\label{Methods}

\subsection*{State-preparation and cooling of ions} 
\begin{figure*}
  \includegraphics[width=\linewidth]{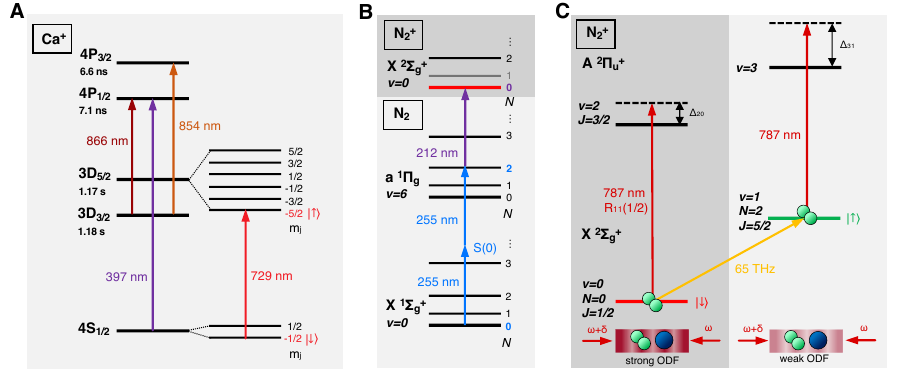}
  \caption{\textbf{Energy levels of Ca$^+$, N$_2$ and N$_2^+$ relevant for the current study.} \textbf{(A)} Energy levels of Ca$^+$ relevant to current experiments (not to scale) and their corresponding radiative lifetimes~\cite{hettrich15a, kreuter05a, meir20a}. The violet and red arrows indicate the transitions used for cooling and coherent manipulation, respectively, while the brown arrows represent repumping transitions. The levels used for the $\text{Ca}^+$ clock transition are indicated in red. \textbf{(B)} Energy-level diagram of the $[2+1']$ resonance-enhanced multiphoton ionisation (REMPI) scheme employed for the rovibrational-ground-state preparation of single N$_2^+$ ions~\cite{shlykov25a}. \textbf{(C)} Optical dipole force (ODF) used for molecular-state readout. Red arrows represent the optical lattice laser around 787~nm detuned from the $R_{11}(1/2)$ component of the $X^2\Sigma_g^+(v''=0)\rightarrow A^2\Pi_u(v'=2)$  vibronic transition in N$_2^+$. When the molecular ion is in the ground state $\ket{\downarrow}_{\text{N}_2^+}$, the optical lattice creates a strong ODF. When the ion is shelved to the rovibrationally excited state $\ket{\uparrow}_{\text{N}_2^+}$ by a pulse from the QCL (yellow arrow), the lattice laser is too far detuned from any spectroscopic transition to create a strong ODF.}        
  \label{fig:diagrams}
\end{figure*}

\noindent Ca$^+$ ions were produced by ionization of neutral Ca atoms in a skimmed atomic beam in the center of the ion trap using a $[1+1']$ resonance-enhanced two-photon scheme with laser beams at 423~nm and 375~nm~\cite{lucas04a}. The Ca$^+$ ions were laser cooled using 397~nm and 866~nm lasers~\cite{meir19a} close to the Doppler limit forming Coulomb crystals in the trap. 
Single N$_2^+$ ions in their rovibrational ground state were produced inside the ion trap by threshold photoionization of internally cold neutral N$_2$ molecules from a molecular beam using a $[2+1']$ REMPI scheme~\cite{shlykov25a}. The single molecular ions were sympathetically cooled into the Ca$^+$ Coulomb crystals~\cite{karl26a, tong10a} which were subsequently reduced to Ca$^+$-N$_2^+$ two-ion strings (inset in Fig.~\ref{fig:schemes}B). The in-phase axial normal mode of the string was then cooled close to its ground state using resolved sideband cooling on the red sideband of the \mbox{3d~$^2$D$_{5/2}(m_j=-5/2)$} $\leftarrow$ \mbox{4s~$^2$S$_{1/2}(m_j=-1/2)$} clock transition of Ca$^+$ at 729~nm~\cite{roguski25a, meir19a}. See Fig.~\ref{fig:diagrams}A,B for the relevant energy levels of the Ca$^+$ ion and of the N$_2$ ionization scheme. 

\subsection*{Generation of a state-dependent optical dipole force and quantum-non-demolition state detection} 
\noindent A state-dependent optical dipole force (ODF) was generated by a 1D traveling optical lattice consisting of two counter-propagating laser beams around 787~nm detuned from the $R_{11}(1/2)$ component of the $X^2\Sigma_g^+(v''=0)\rightarrow A^2\Pi_u(v'=2)$ vibronic transition in N$_2^+$~\cite{meir19a, sinhal20a}. The two laser beams were detuned from each other by the frequency of the in-phase axial motional mode of the N$_2^+$-Ca$^+$ ion string in the trap (674 kHz in the current experiments). Under these conditions, a strong ODF was generated for N$_2^+$ in the $X^2\Sigma_g^+(v=0, N=0)$ rovibrational ground state resonantly exciting the in-phase axial motion of the ions~\cite{meir19a,sinhal20a,najafian20a}. For other rovibrational states of the molecular ion ($X^2\Sigma_g^+(v\geq0,N\geq2)$), the frequency of the optical lattice was too far detuned from any spectroscopic transition to produce a strong ODF and, therefore, no motional excitation ensued~\cite{sinhal20a} (Fig.~\ref{fig:diagrams}C).
Motional excitation of the two-ion strings was detected by Rabi sideband thermometry on the Ca$^+$ clock transition as detailed in~\cite{sinhal20a, roguski25a}.

\subsection*{Frequency stabilization of the spectroscopy lasers}
\noindent The frequency-stabilization and laser-beam-delivery scheme is illustrated in Fig.~\ref{fig:schemes}B. The repetition rate of the optical frequency comb (OFC) was phase-locked to the 729~nm external-cavity diode laser (ECDL) addressing the Ca$^+$ clock transition which was itself stabilized to a high-finesse ultralow expansion cavity achieving a linewidth at the sub-kilohertz level. 
Additionally, the OFC was referenced to an ultra-stable 1572~nm frequency signal disseminated by the Swiss Federal Institute of Metrology (METAS) to ensure long-term stability and traceability to the SI second realized by the Swiss primary frequency standard~\cite{husmann21a}. The quantum cascade laser (QCL) operating at 4574~$\mu$m was locked to the OFC at a wavelength of 629~nm following sum-frequency generation (SFG) with 729~nm laser radiation. The beat note between the 629~nm light and the closest comb tooth was detected using a balanced photodetector and stabilized to a reference signal from a direct digital synthesizer (DDS)~\cite{sinhal23a}. By tuning the DDS reference frequency, coarse scanning of the locked QCL frequency was achieved. Fine scanning of the QCL was implemented by an acousto-optic modulator using a linear ramp around its center frequency. 

\section*{Acknowledgments}
\noindent We thank Georg Holderried, Dominik Baumgartner, Dr. Anatoly Johnson, Grischa Martin and Philipp Knöpfel for technical support as well as Prof. Piet O. Schmidt and Dr. Nicol\'as A. Nu\~nez Barreto for useful discussions. 

\subsection*{Funding}
\noindent This work was supported by the Swiss National Science Foundation, grant nrs. 200021\_204123, TMAG-2\_209193, the European Partnership on Metrology (Funder ID: 10.13039/100019599, grant nr. 23FUN04 COMOMET) and the University of Basel. 

\subsection*{Author contributions}
\noindent A.S. and S.W. developed the experimental methodology.
M.D., A.S., R.K. and U.J. carried out the experiments and analyzed the data.
A.S., M.R., R.K. and M.D., contributed to the development and characterization of the experimental setup.
S.W. conceived and supervised the project.
A.S., M.D. and S.W. drafted the paper, all authors contributed to the final version.

\subsection*{Data availability}
\noindent The primary data supporting the findings of this study are available on Zenodo \textit{https://doi.org/10.5281/zenodo.18935655}.


\section*{Supplementary Information}

\subsection*{Effective Hamiltonian}
\label{sec:hamiltonian}

\noindent The theoretical line positions were calculated by numerical diagonalization of the effective spin-rovibrational-Zeeman Hamiltonian of the electronic ground state $X~^2\Sigma_g^+$ of N$_2^+$~\cite{berrahmansour91a, balasubramanian94a, brown03a, najafian20b},
\begin{equation}
\label{N2_hamiltonian}
\hat{H}=\hat{H}_{vib}+\hat{H}_{rot}+\hat{H}_{fs}+\hat{H}_{hfs}+\hat{H}_{Z}.
\end{equation}
Here, the first three terms on the right-hand side describing vibrations, rotations and fine (i.e., spin-rotation) structure are diagonal in a Hund's case $b_{\beta_J}$ basis~\cite{berrahmansour91a}. Their eigenvalues are given by,
\begin{equation}
\label{H_vib}
\hat{H}_{vib}=G_v,  
\end{equation}

\begin{equation}
\label{H_rot}
\hat{H}_{rot}=B_v N(N + 1) - D_v (N(N + 1))^2, 
\end{equation}

\begin{equation}
\label{H_fs}
\hat{H}_{fs}=\gamma_{v,N}(J(J+1)-N(N+1)-S(S+1))/2.
\end{equation}

\noindent
Here, $ G_v,~B_v, ~D_v$ are the vibrational, rotational and centrifugal-distortion constants in a specific vibrational level $v$ of the electronic ground state. $\gamma_{v,N}$ is the electron spin-rotation coupling constant, which includes a centrifugal-correction term $\gamma_{v,N}=\gamma_{v}+\gamma_{Nv}(N(N+1))$~\cite{berrahmansour91a}. 

The effective hyperfine Hamiltonian is given by~\cite{berrahmansour91a},
\begin{equation}
\label{H_hfs}
\hat{H}_{hfs}=\hat{H}_{bF}+\hat{H}_{t}+\hat{H}_{eqQ}+\hat{H}_{cI},
\end{equation}
where $\hat{H}_{bF}$ represents the Fermi-contact interaction, $\hat{H}_{t}$ is the dipolar hyperfine interaction, $\hat{H}_{eqQ}$ is the electric-quadrupole hyperfine interaction, and $\hat{H}_{cI}$ is the magnetic nuclear spin-rotation interaction. The matrix elements for these four terms have been derived in~\cite{berrahmansour91a} and can also be found in~\cite{najafian20b}. The values of the relevant hyperfine spectroscopic constants reported in literature~\cite{scholl98a, berrahmansour91a} are listed in Table~\ref{tab:N2_plus_constants_hfs}.

The effective Zeeman Hamiltonian $\hat{H}_{Z}$ includes four first-order contributions from the interaction of an external magnetic field $\vb{B}$ with the magnetic dipole moments arising from the electron spin $\vb{S}$, the molecular rotation $\vb{N}$ and the total nuclear spin $\vb{I}$~\cite{brown03a, brown78a, ma09a, chen06a},
\begin{equation}
\label{H_z}
\begin{aligned}
\hat{H}_{Z}
&=g_s \mu_B T^{1}_{p=0}(\mathbf{B})T^{1}_{p=0}(\mathbf{S}) \\
&-g_r \mu_B T^{1}_{p=0}(\mathbf{B})T^{1}_{p=0}(\mathbf{N}) \\
&-g_n \mu_N T^{1}_{p=0}(\mathbf{B})T^{1}_{p=0}(\mathbf{I}) \\
&+g_l \mu_B T^{1}_{p=0}(\mathbf{B})\sum_{q=\pm 1}\mathcal{D}^{1}_{p=0,q}(\omega)^{*}T^{1}_{q}(\mathbf{S}).
\end{aligned}
\end{equation}

\noindent Here, $g_s$, $g_r$, $g_n$ are the $g$-factors for the electron spin, rotation, nuclear spin and $\mu_B$ ($\mu_N$) is the Bohr (nuclear) magneton. The last term in $\hat{H}_{Z}$ represents the anisotropic correction to the electron-spin Zeeman interaction, and $g_l$ is the corresponding effective $g$-factor (see Table~\ref{tab:N2_plus_constants_rest}). $T_p^{1}$ denotes a rank~1 spherical tensor operator in the space-fixed coordinate system (subscript $p$), $\mathcal{D}^{1}_{pq}(\omega)$ is a Wigner rotation-matrix element and the subscript $q$ denotes spherical tensor components in the molecule-fixed coordinate system. The $p=0$ component of the space-fixed coordinate system is taken to be aligned with the direction of the magnetic field oriented along the $z$ axis, $T^{1}_{p=0}(\mathbf{B}) = B_z \hat{z}$. The diagonal terms of the interaction of the magnetic field with the electronic orbital angular momentum $\mathbf{L}$ vanish in a $\Sigma$ electronic state and terms of higher order in the magnetic field are neglected~\cite{brown03a}.
The matrix elements for these four terms can be found in~\cite{najafian20b}.

\subsection*{Assignment of observed spectroscopic transitions}

\noindent The spectral features observed in Fig.~\ref{fig:main_spectrum} were assigned based on the calculation of the (hyper)fine structure of the S(0) rovibrational line. Because the observed transitions are all part of the same rotational component of the infrared fundamental, the rovibrational energy can be treated as a common offset for the calculation of the splitting between the lines, and only the fine (Equation~\ref{H_fs}) and hyperfine (Equation~\ref{H_hfs}) parts of the Hamiltonian were considered for that purpose.

Table~\ref{tab:lines_splitting} lists the spectroscopic lines \(L1, \ldots, L6\) observed in Fig.~\ref{fig:main_spectrum} and compares the experimental line splittings with the theoretical ones calculated by diagonalizing the effective hyper(fine) Hamiltonian with the spectroscopic constants from~\cite{scholl98a,berrahmansour91a} (see Table~\ref{tab:N2_plus_constants_hfs}). The separations between the lines observed in this study are consistent with the calculated values within an experimental uncertainty of 1$\sigma$. This enables an unambiguous assignment of all lines observed in the experimental spectrum as indicated in Fig.~\ref{fig:main_spectrum} and Table~\ref{tab:extracted_lines}.

\subsection*{Determination of the fundamental vibrational frequency of N$_2^+$ from the spectroscopic data}
\label{sec:deltag}

\noindent From the spectroscopic data obtained in the current study as presented in Fig.~\ref{fig:main_spectrum} and Table~\ref{tab:extracted_lines}, an improved value of the fundamental vibrational frequency $\Delta G_{10}=G_1-G_0$ of N$_2^+$ can be obtained. As for all lines of the observed S(0) rovibrational manifold the vibrational and rotational constants are identical, $\Delta G_{10}$ is most precisely determined from the frequency $f_{L4}$ of the $L4$~$(\ket{J''=1/2} \rightarrow \ket{J'=5/2})$ transition in the $I=0$ isomer which is not affected by hyperfine structure. Based on the effective Hamiltonian, $\Delta G_{10}$ is computed from $f_{L4}$ as, 
\begin{equation}
    \Delta G_{10}=f_{L4}-6B_1+36D_1-\gamma_1-6\gamma_{N1}, \label{eq:deltag}
\end{equation}
where $B_1, D_1$ as well as $\gamma_{1}$ and $\gamma_{N1}$ are the rotational, centrifugal-distortion and spin-rotation constants of the $v=1$ level. Note that the corresponding parameters of the $v=0$ level do not appear in Equation~(\ref{eq:deltag}) as the transition starts from the rotational ground state $N=0$.

Because in previous studies the (spin)rotational constants have been determined with much greater precision than $\Delta G_{10}$, Equation~\ref{eq:deltag} was evaluated using mean values of these parameters reported in the literature. Weighted means of the literature values of $B_1,~D_1$ and $\gamma_{1}$ (see Tab.~\ref{tab:mean_constants}) were calculated according to,
\begin{equation}
    \label{weighted_mean}
    \overline{C}=\frac{\sum_i{w_iC_i}}{\sum_i w_i},
\end{equation}
where $C_i$ are the relevant spectroscopic constants with uncertainties $\delta C_i$, $w_i=1/\delta C_i^2$ are their weights and 
\begin{equation}
\label{weighted_mean_error}
\delta \overline{C}=\frac{1}{\sqrt{\sum_iw_i}}   
\end{equation}
is the weighted uncertainty of the mean value.

\begin{figure*}
    \includegraphics[width=\linewidth]{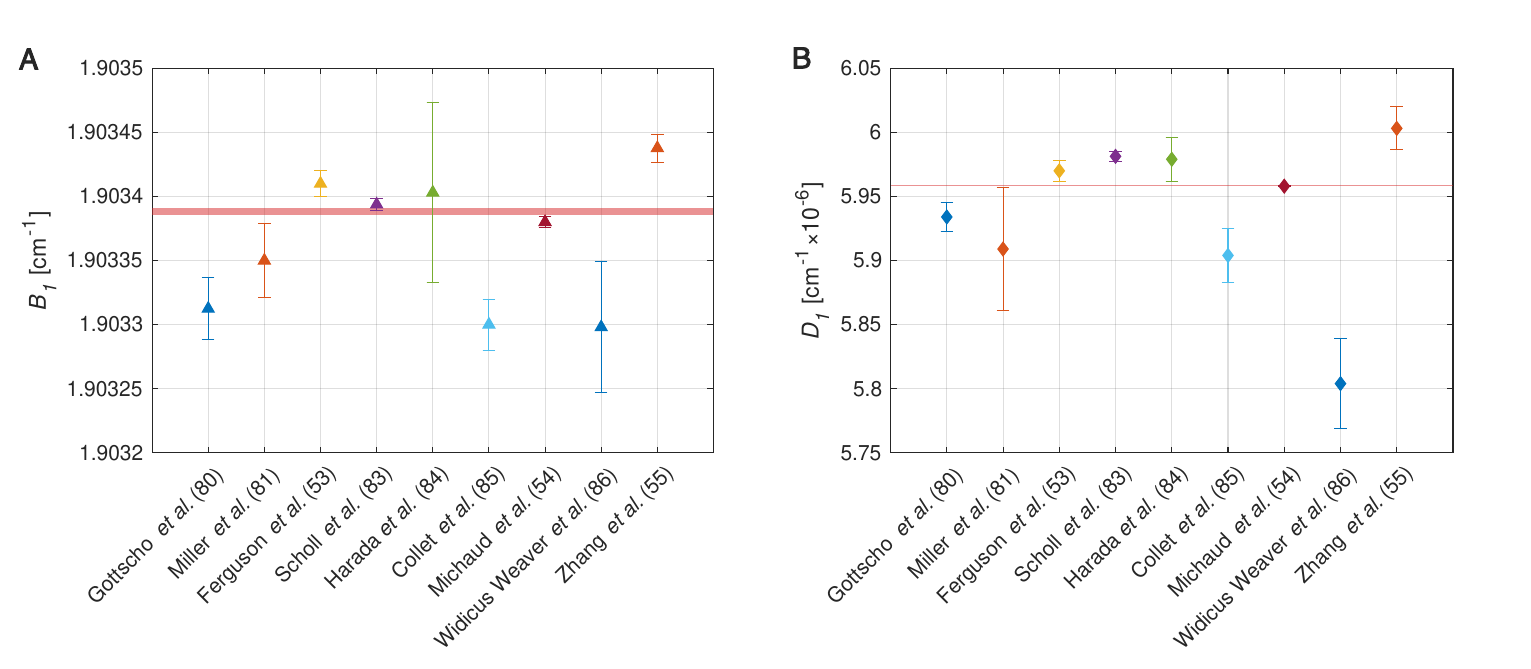}
    \caption{\textbf{Literature values of rotational spectroscopic constants.} Values of the \textbf{(A)} rotational ($B_1$) and \textbf{(B)} centrifugal-distortion ($D_1$) constants of the $v=1$ vibrational state of N$_2^+$ reported in the literature. The red shaded areas indicate the uncertainty of the weighted mean of the literature values. Numerical values are given in Tab.~\ref{tab:mean_constants}}
    \label{fig:mean_constants}
\end{figure*}

The mean values $\bar{B}_1$ and $\bar{D}_1$ and their uncertainties are plotted in Fig.~\ref{fig:mean_constants} together with the relevant data from the literature. The mean value of the fine-structure constant $\gamma_1$ is dominated by the value reported in Ref.~\cite{berrahmansour91a} due to its significantly lower uncertainty compared to those reported in the other studies. Therefore, $\gamma_1$ and $\gamma_{N_1}$ were adopted straight from~\cite{berrahmansour91a}.

Thus, replacing the rotational constants $B_1,D_1$ in Eq.~(\ref{eq:deltag}) by the mean values $\bar{B}_1, \bar{D}_1$ from the literature, $\Delta G_{10}$ was obtained as $\Delta G_{10}=65197356.2(21)$~MHz or $2174.74971(7)$~cm$^{-1}$. The uncertainty quoted for $\Delta G_{10}$ was calculated by propagating the experimental uncertainty $\delta f_{L4}$ of the position of the $L4$ line and the uncertainties of the mean spectroscopic constants from Table~\ref{tab:mean_constants} through Equation~(\ref{eq:deltag}) as,
\begin{equation}
    \label{eq:vib_const_err}
    \delta \Delta G_{10}=\sqrt{\delta f_{L_4}^2+(6\delta \bar{B}_1)^2+(36\delta \bar{D}_1)^2+\delta \gamma_1^2+(6\delta\gamma_{N1})^2}.
\end{equation}

\subsection*{Fit of the hyperfine spectrum and simulation of the spectrum}
\label{sec:hfsfit}

\noindent To obtain updated values of the hyperfine constants of N$_2^+$, calculated line positions derived from the effective Hamiltonian Equation~\ref{N2_hamiltonian} (without the Zeeman contribution $H_Z$) were fitted to the spectrum in Fig.~\ref{fig:main_spectrum} using a least-squares procedure, where the hyperfine constants $b_{F0}, b_{F1}$ and $t_1$ were treated and previously reported values, listed in Tab.~\ref{tab:N2_plus_constants_hfs}, were used as initial guess. The (spin)rovibrational constants were fixed to the values $\bar{B}_1, \bar{D}_1, \gamma_1$ and $\Delta G_{10}$. The other hyperfine spectroscopic constants were fixed to the values listed in Tab.~\ref{tab:N2_plus_constants_hfs}, because of their negligible impact on the line positions. The results of the fit and the hyperfine constants obtained are given in Tables~\ref{tab:hf_fit} and \ref{tab:hf_constants_fitted} respectively, and the spectrum based on the obtained constants is plotted in the bottom of Fig.~\ref{fig:main_spectrum}. The strength of the spectroscopic lines were calculated following~\cite{najafian20b}.

\subsection*{Calculation of uncertainties of the line positions} 
\label{sec:errors}

\noindent The uncertainties of the observed spectroscopic transitions shown in Fig.~\ref{fig:main_spectrum} as listed in Table~\ref{tab:extracted_lines} were calculated according to,
\begin{equation}
    \label{eq:exp_err}
    \delta f_{exp}=\sqrt{\Delta f^2+\delta_{fit}^2+\delta f_{QCL}^2+\delta f_{B}^2},
\end{equation}
where $\Delta f=2$~MHz is the sweep width of the mid-IR QCL, $\delta_{fit}$ corresponds to the Gaussian fit error for determining the center of adjacent signals (ranging from 30 to 130~kHz), $\delta f_{QCL}\simeq2$~kHz is the conservatively estimated uncertainty in the frequency of the QCL~\cite{sinhal23a} and $\delta f_{B}$ is the estimated uncertainty in the transition frequency due to the magnetic field fluctuations of about 2 mG~\cite{najafian20b} which is negligible in comparison with other sources of errors.

\subsection*{Calculation of the Rabi frequency of electric-quadrupole vibrational transitions in the electronic ground state of N$_2^{+}$}

\noindent The Rabi frequency of an electric-quadrupole transition is given by~\cite{james98a},
\begin{equation}
\label{Rabi_E2}
\Omega^{E2}=\frac{E\omega}{2\hbar c}
\sum_{\alpha, \beta}
\abs{
\bra{\phi_j}q\hat{r}_{\alpha}\hat{r}_{\beta}\ket{\phi_i}
\varepsilon_{\alpha\beta}},
\end{equation}
where $E$ and $\omega$ are the amplitude and frequency of the electric field of the plane electromagnetic wave, $q\hat{r}_{\alpha}\hat{r}_{\beta} $ is the electric-quadrupole moment operator, $q$ is the elementary charge, $\varepsilon_{\alpha\beta}=\epsilon_{\alpha}\kappa_{\beta}$ represents the polarization–propagation second-rank tensor associated with the electric-field gradient of a plane wave, where $\epsilon_{\alpha}$ are the components of polarization vector and $\kappa_{\beta}$ are the components of the unit vector in the direction of wave propagation, $\hbar$ is the reduced Planck constant and $c$ is the speed of light.

In spherical tensor notation~\cite{zare88a}, Equation~(\ref{Rabi_E2}) can be reformulated as,
\begin{equation}
\label{E2_moment_spher}
\Omega^{E2}=\frac{E\omega}{2\hbar c}
\sum_{k, p}
\abs{(-1)^{k-p}
\bra{\phi_j}(T^{(k)}_{-p}(\vu{Q}))\ket{\phi_i}
T^{(k)}_p(\vb{\varepsilon})},
\end{equation}
where $k=0,1,2$, $p=-k,...,k$, tensor $\vb{\varepsilon}$ is the electric-field gradient tensor with Cartesian components $\varepsilon_{\alpha\beta}$ as above and $\vu{Q}$ is the electric-quadruple-moment tensor operator whose Cartesian components are usually written in traceless form~\cite{bunker98a} as,
\begin{equation}
\label{quadrupole_operator}
\hat{Q}_{\alpha\beta}:=q(\hat{r}_{\alpha}\hat{r}_{\beta}-\delta_{\alpha\beta}\frac{\hat{r}^2}{3}),
\end{equation}
where $\hat{r}^2=\hat{r}^2_x+\hat{r}^2_y+\hat{r}^2_z$.

As the quadrupole tensor $\vu{Q}$ is traceless ($\sum_{\alpha}\hat{Q}_{\alpha\alpha}=0$) and symmetric ($\hat{Q}_{\alpha\beta}=\hat{Q}_{\beta\alpha}$), all components of the sum in Equation~(\ref{E2_moment_spher}) except for $k=2$ are zero: $T^{(0)}_{0}(\vu{Q})=0$, $T^{(1)}_{0,\pm1}(\vu{Q})=0$~\cite{zare88a, brown03a}. Therefore,
\begin{equation}
\label{Rabi_E2_2}
\Omega^{E2}=\frac{E\omega}{2\hbar c}
\sum_{p=-2}^{2}\abs{(-1)^{2-p}\bra{\phi_j}(T^{(2)}_{-p}(\vu{Q}))\ket{\phi_i}T^{(2)}_p(\vb{\varepsilon})}.
\end{equation}

The Cartesian components of the second rank spherical tensor $T^{(2)}_p(\vb{\varepsilon})$ are given by~\cite{zare88a, brown03a},
\begin{equation}
    \begin{aligned}
        \label{spher_tens_cart_comp_second}
        T^{(2)}_{0}(\vb{\varepsilon})&=\frac{1}{\sqrt{6}}[2\varepsilon_{zz}-\varepsilon_{xx}-\varepsilon_{yy}],\\
        T^{(2)}_{\pm1}(\vb{\varepsilon})&=\mp\frac{1}{2}[(\varepsilon_{xz}+\varepsilon_{zx})\pm i(\varepsilon_{yz}+\varepsilon_{zy})],\\
        T^{(2)}_{\pm2}(\vb{\varepsilon})&=\frac{1}{2}[(\varepsilon_{xx}-\varepsilon_{yy})\pm i(\varepsilon_{xy}+\varepsilon_{yx})].
    \end{aligned}
\end{equation}
Assuming, as in the present experiments, a linearly polarized laser beam with polarization and propagation direction perpendicular to the quantization axis defined by the magnetic field $B$ ($\vb{k}\perp\vb{\epsilon}\perp\vb{B}$) such that $\vb{\epsilon}=(0,1,0)$ and $\vb{\kappa}=(1,0,0)$, only transitions with $p=\pm2$ ($\Delta m_F=\pm2)$ are driven and one obtains,
\begin{equation}
\label{T2}
T^{(2)}_{\pm2}(\vb{\varepsilon})=\pm \frac{i}{2},
\end{equation}
and therefore
\begin{equation}
\label{Rabi_E2_2_p2}
\Omega^{E2,p=\pm2}=\frac{1}{2}\frac{E_0\omega}{2\hbar c}\bra{\phi_j}(T^{(2)}_{\pm2}(\vu{Q}))\ket{\phi_i}.
\end{equation}
\newline
$S_{ij}=\abs{\bra{\phi_j}(T^{(2)}_{\pm2}(\vu{Q}))\ket{\phi_i}}^2$ is then the line strength of the transition as defined in~\cite{najafian20b}. 
\begin{widetext}
\noindent The electric-quadrupole-transition matrix element can be separated into angular ($\mathcal{A})$ and radial ($\mathcal{R}$) parts~\cite{najafian20b, karl67a},

\begin{equation}
\bra{\phi_j}(T^{(2)}_{\pm2}(\vu{Q}))\ket{\phi_i}
=\mathcal{A}(\dots, F_j,m_{Fj}, F_i,m_{Fi},\pm2)\mathcal{R}(v_1,v_0)\\
=\mathcal{A}(\dots, F_j,m_{Fj},F_i,m_{Fi},\pm2)
\sqrt{\frac{3}{2}}\frac{dQ^{(m)}_{zz}}{dR}R_e\sqrt{\frac{B_e}{\omega_e}}.
\end{equation}

\noindent

\noindent
Here $Q^{(m)}_{zz}$ denotes the traceless electric quadrupole component in the molecular (body-fixed) frame~\cite{papousek89a}, $R$ is the internuclear separation and $R_e$ is the equilibrium bond length. The constants $B_e$ and $\omega_e$ are the equilibrium rotational and vibrational constants. Since we are unaware of any experimental data for the derivative of the electric quadrupole moment in N$_2^+$, we used a theoretical value derived from~\cite{bruna04a}. In~\cite{bruna04a}, $Q_{zz}$ was defined as being proportional to $(3z^{2}-r^{2})$, whereas in Equation~(\ref{quadrupole_operator}) we use $(z^{2}-r^{2}/3)$ (see also Appendix~A.1 and B of~\cite{germann16c} and~\cite{lawson97a}). Therefore, the reported values of $Q^{(m)}_{zz}$ and $dQ^{(m)}_{zz}/dR$ depend on the definition of the quadrupole operator. Consequently,
\begin{equation}
\frac{dQ_{zz}}{dR}^{(m,~\text{this work)}}=\frac{2}{3}\frac{dQ_{zz}}{dR}^{(m,~\text{Ref. \cite{bruna04a}})}=\frac{2}{3}\times2.63(3)~ea_0.
\end{equation}

\noindent
Finally, for quadrupole transitions from $v=0$ to $v=1$ in the ground electronic state of N$_2^+$ driven by a linearly polarized laser beam with $\vb{k}\perp\vb{\epsilon}\perp\vb{B}$ (yielding $p=\pm2$), the Rabi frequency is obtained as:

\begin{equation}
\label{Rabi_E2_v01}
\Omega^{E2,p=\pm2}
=
\frac{1}{2}\frac{E_0\omega}{2\hbar c}\,
\mathcal{A}(\dots,F_j,m_{Fj},F_i,m_{Fi},\pm2)\,
\sqrt{\frac{3}{2}}\,
\left(\frac{2}{3}\frac{dQ_{zz}}{dR}^{(m,~\text{Ref. \cite{bruna04a}}}\right)
R_e\sqrt{\frac{B_e}{\omega_e}} .
\end{equation}

\end{widetext}

\subsection*{Calculation of population-transfer probabilities using RAP}

\noindent In the electronic ground state of N$_2^+$, two Zeeman sublevels $\ket{v''=0, N''=0, J'',F'',m_F''}$ and\\ $\ket{v'=1, N'=2, J',F',m_F'}$ coupled by a laser with frequency $\omega$ were simulated as an effective two-level system within the rotating-wave approximation with resonance frequency $\omega_{eg}$ using the optical Bloch equations (OBE)~\cite{foot05a},
\begin{equation}
\label{OBEs}
    \begin{aligned}
    \dot{\rho}_{ee}&=\frac{i}{2}\Omega(\rho_{eg}-\rho_{ge})-\Gamma_p \rho_{ee},\\
    \dot{\rho}_{gg}&=-\frac{i}{2}\Omega(\rho_{eg}-\rho_{ge})+\Gamma_p \rho_{ee},\\
    \dot{\rho}_{eg}&=\frac{i}{2}\Omega(\rho_{ee}-\rho_{gg})-(\Gamma_c-i\Delta) \rho_{eg},\\
    \dot{\rho}_{ge}&=-\frac{i}{2}\Omega(\rho_{ee}-\rho_{gg})-(\Gamma_c+i\Delta) \rho_{ge},
    \end{aligned}   
\end{equation}
where $\rho_{ee}$ and $\rho_{gg}$ are the populations of the excited and ground state of the two-level system, respectively, $\rho_{eg}$ and $\rho_{ge}$ are the coherences, $\Omega$ is the Rabi frequency, $\Delta=\omega_{eg}-\omega$ is the laser detuning, and $\Gamma_p$ and $\Gamma_c$ are effective population-decay and decoherence rates, respectively.

\begin{figure*}
    \includegraphics[width=\linewidth]{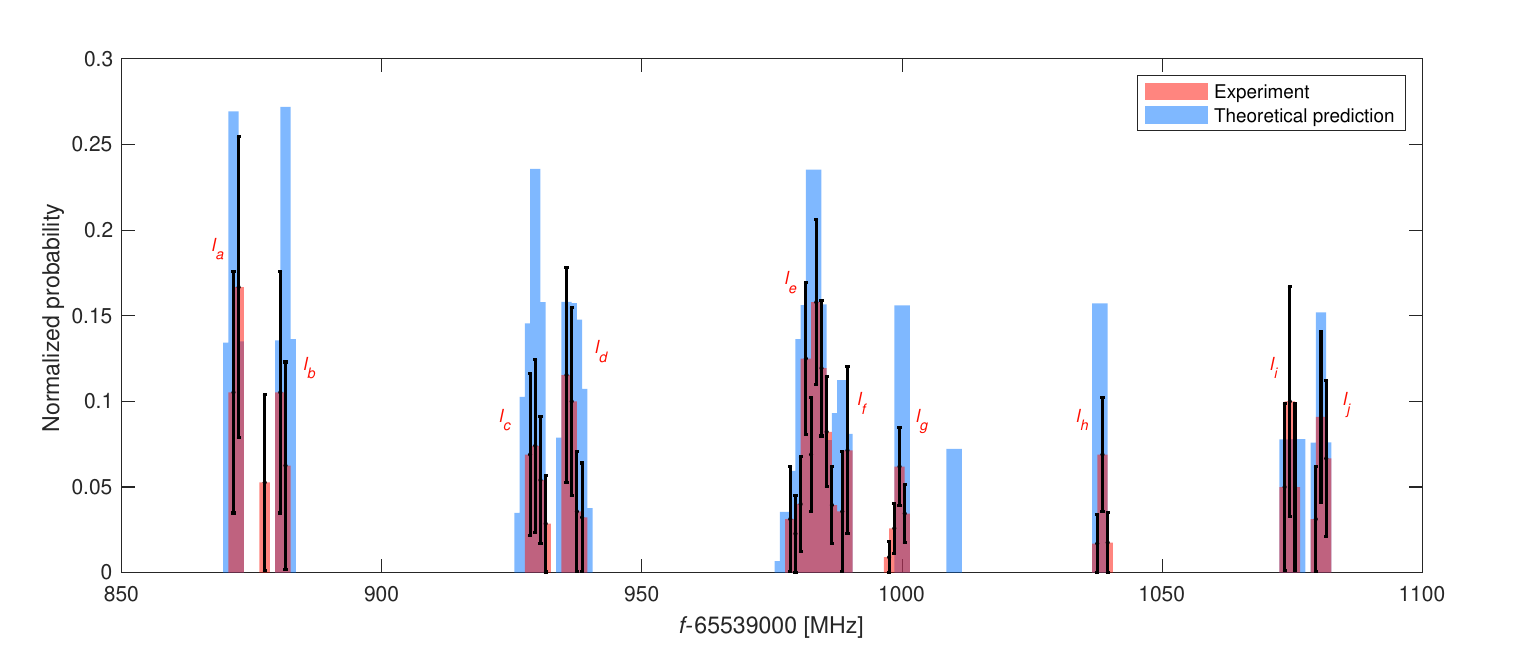}
    \caption{\textbf{Normalized population-transfer probabilities.} Simulation (blue) of the experimentally observed population-transfer probabilities (red) as shown in Fig.~\ref{fig:main_spectrum} of the main text. Error bars indicate binomial uncertainties of the experimental data. See Supplementary Information for details.}
    \label{fig:rapspectrumsimulation}
\end{figure*}

To simulate the two-level system interacting with a light pulse of frequency $\omega(t)$ swept over the resonance frequency $\omega_{eg}$, the OBE (\ref{OBEs}) were numerically solved assuming a linear chirp,
\begin{equation}
\label{detuning}
\Delta(t)=-\frac{\Delta\omega}{2}+\Delta\omega\frac{t}{t_{\mathrm{chirp}}},
\end{equation}
where $t\in [0, t_{chirp}]$ and the chirped pulse crosses the resonance at half duration, such that $\Delta(t_{chirp}/2)=0.$ 

The results of the simulation for the two different chirp time intervals, $t_{chirp}$ = 500 ms and $t_{chirp}$ = 1500 ms, for the same sweep width $\Delta\omega=2\pi\Delta f=2\pi\times$2 MHz are shown in Fig.~\ref{fig:RAP} of the main text. The Rabi frequencies of the individual Zeeman components of (hyper)fine transitions were obtained from the calculated strength of the transitions and the laser intensity employed experimentally (see the simulated spectrum in the bottom panel of Fig.~\ref{fig:main_spectrum} in the main text).

The top panel of Fig.~\ref{fig:main_spectrum} of the main text shows a histogram of the normalized experimental probabilities of a successful population transfer by RAP within each frequency bin. 
Assuming equal probabilities for populating the relevant Zeeman states upon photoionization and taking to account that from a particular Zeeman level $i$, there is only one allowed transition to an excited Zeeman state $j$ within the experimentally employed 2 MHz RAP interval and magnetic field 4.7 Gauss, expected normalized experimental probabilities $P_{EXP}^{bin}$ could be calculated as:
\begin{equation}
    P_{EXP}^{bin}=\frac{\sum_{i}P_{RAP,ij}^{bin}}{12}.
\end{equation}
Here, the RAP probabilities $P_{RAP,ij}^{bin}$ were calculated using a Landau-Zener model of adiabatic transitions~\cite{landau32a,zener32a,camparo84a}, incorporating an experimentally estimated effective decoherence time $T_2\simeq25$~ms and the calculated Rabi frequencies as shown in the bottom panel of Fig.~\ref{fig:main_spectrum} of the main text:
\begin{equation}
    \label{RAP_probability}
    P_{RAP,ij}^{bin}=\left[1-\exp\left(-\frac{\pi\Omega_{ij}^2}{2\abs{\alpha}}\right)\right]\times\left[1-\frac{3\pi\Omega_{ij}}{8\abs{\alpha}T_2}\right],
\end{equation}
where $\alpha=\Delta\omega/t_{chirp}$ is the sweep rate.

A comparison of the theoretical prediction with the experimental data, including binomial error bars, is shown in Fig.~\ref{fig:rapspectrumsimulation}. The agreement between theory and experiment is satisfactory considering the uncertainty in the theoretical Rabi frequencies. The single signal at 877.5 MHz is most likely an outlier due to a calibration offset during this specific measurement. 


\begin{table*}
\centering
\caption{\textbf{(Hyper)fine spectroscopic constants of $^{14}$N$_2^+$} in the $v = 0$ and $v = 1$ vibrational states of the electronic ground state $X^2\Sigma_g^+$ used for the calculation of (hyper)fine splittings. The numbers in parentheses indicate the uncertainties quoted in the literature.}
    \label{tab:N2_plus_constants_hfs}
    \begin{tabular}{|l||l|c||l|c|}
    \hline  
    Constant& $v = 0$& Ref & $v = 1$ & Ref \\
    \hline
    \hline  
    $\gamma_v$ [MHz] & 280.25(45) & \cite{scholl98a} & 276.92253(13) & \cite{berrahmansour91a} \\
    $\gamma_{Nv}$ [kHz] & 0 & \cite{scholl98a} & -0.39790(23) &\cite{berrahmansour91a} \\
    $b_{Fv}$ [MHz] & 102.4(11) & \cite{scholl98a} & 100.6040(15) & \cite{berrahmansour91a} \\
    $t_v$ [MHz] & 23.3(10) & \cite{scholl98a} & 28.1946(13) & \cite{berrahmansour91a} \\
    $t_{Nv}$ [Hz] & 0 & \cite{scholl98a} & -73.5(27) & \cite{berrahmansour91a} \\
    $eqQ_v$ [MHz] & -& & 0.7079(60) & \cite{berrahmansour91a} \\
    $c_{Iv}$ [kHz] & -& & 11.32(85) & \cite{berrahmansour91a}\\
    \hline
\end{tabular}  
\end{table*}

\begin{table*}
\centering
\caption{\textbf{Effective values of $g$-factors in the effective Zeeman Hamiltonian $\hat{H}_Z$} of the $v = 0$ and $v = 1$ vibrational states of the electronic ground state $X^2\Sigma_g^+$.}
    \label{tab:N2_plus_constants_rest}
    \begin{tabular}{|l||l|c||l|c|}
    \hline
    Constant& $v = 0$ & Ref& $v = 1$ & Ref \\
    \hline
    \hline
    $g_s \mu_B$ [MHz/G] & 2.8025 & \cite{kajita15a} & 2.8025 & \cite{kajita15a} \\
    $g_r \mu_B$ [Hz/G] & 50.107 & \cite{kajita15a} & 49.547 & \cite{kajita15a} \\
    $g_n \mu_N$ [Hz/G] & 307.92 & \cite{kajita15a} & 307.92 & \cite{kajita15a} \\
    $g_l \mu_B$ [Hz/G] & -3793 & \cite{bruna04a} & -3821 & \cite{bruna04a} \\
    \hline
\end{tabular}  
\end{table*}

\begin{table*}
\centering
\caption{\textbf{Observed and calculated splittings between the transitions $Li, i=1,\dots,6$.} The errors quoted for experimental values correspond to the 1$\sigma$ experimental uncertainty. The errors of the calculated values were propagated from the errors of the spectroscopic constants in the literature. All values are given in MHz.}
\footnotesize
\begin{tabular}{|c|c||c|c||c|c||c|c|}
\hline
Line $Li$ & $f_0-f'$ &$(Li-L3)_{exp}$ & $(Li-L3)_{calc}$ & $(Li-L4)_{exp}$ & $(Li-L4)_{calc}$ & $(Li-L5)_{exp}$ & $(Li-L5)_{calc}$ \\
\hline \hline
$L1$ & 876.5(14) &  -106.9(25) & -106.6115(46) & -123.1(25)& -124.6(11) & -162.0(25) & -165.4(20) \\
$L2$ & 932.8(14) &  -50.6(25) & -50.6034(42) &-66.8(25) &-68.6(11) & -105.7(25) & -109.4(20) \\
$L3$  & 983.4(20) & - & - & -16.2(28) & -18.0(11) & -55.1(28) & -58.8(20) \\
$L4$  & 999.6(20)  &  16.2(28) & 18.0(11) & - & - & -38.9(28) & -40.8(17)\\
$L5$ & 1038.5(20)&  55.1(28) & 58.8(20) &38.9(28) & 40.8(17)& - & - \\
$L6$  & 1077.6(14) &   94.3(25) & 97.1(20)& 78.1(25)&79.2(17)& 39.1(25) &38.3635(72) \\
\hline
\end{tabular}
\label{tab:lines_splitting}
\end{table*}

\begin{table*}
\centering
\caption{\textbf{Center frequencies $f_0$ of the observed spectral features in Fig.~\ref{fig:main_spectrum} of the main text and their assignments.} Values are relative to $f'=65539000$~MHz.}
\footnotesize
\begin{tabular}{|c||l|l|}
\hline
Observed line & Assignment &  $f_0-f'~\rm{[MHz]}$ \\
\hline \hline
$l_a$ & Zeeman components of $L1$ & 872.2(20)\\
$(l_a+l_b)/2$ & $L1. \ket{F''=5/2} \rightarrow \ket{F'=5/2}$ & 876.5(14)  \\
$l_b$ & Zeeman components of $L1$ & 880.9(20) \\
$l_c$  & Zeeman components of $L2$ & 929.5(20)\\
$(l_c+l_d)/2$  & $L2. \ket{F''=5/2} \rightarrow \ket{F'=7/2}$ & 932.8(14)  \\
$l_d$  & Zeeman components of $L2$ & 936.0(20)\\
$l_e$  & $L3. \ket{F''=5/2} \rightarrow \ket{F'=9/2}$ & 983.4(20)  \\
$l_f$  & $\ket{J''=1/2, m_F=1/2} \rightarrow \ket{J'=5/2,m_F=-3/2}$ & 989.2(20) \\
$l_g$  & $L4. \ket{J''=1/2, m_F=\pm1/2} \rightarrow \ket{J'=5/2,m_F=\pm5/2}$ & 999.6(20) \\
$l_h$  & $L5. \ket{F''=3/2} \rightarrow \ket{F'=1/2}$ & 1038.5(20) \\
$l_i$  & Zeeman components of $L6$ & 1074.5(20)\\
$(l_i+l_j)/2$ & $L6. \ket{F''=3/2} \rightarrow \ket{F'=3/2}$ & 1077.6(14)  \\
$l_j$  & Zeeman components of $L6$ & 1080.8(20)\\
\hline
\end{tabular}
\label{tab:extracted_lines}
\end{table*}
\begin{table*}
    \centering
    \caption{\textbf{(Spin)rotational spectroscopic constants reported in the literature.} Values of the rotational ($B_1$), centrifugal distortion ($D_1$) and fine-structure ($\gamma_1$) constants for the $v=1$ state of N$_2^+$ reported in the literature and their mean values calculated according to Eqs.~(\ref{weighted_mean},\ref{weighted_mean_error}).}
    \begin{tabular}{|c|c|c||l|l|l|}
    \hline
        Authors & Year & Ref & $B_1$ [cm$^{-1}$] & $D_1$ [cm$^{-1}\times 10^{-6}$]  & $\gamma_1$~[MHz] \\
        \hline
        \hline
        Gottscho \emph{et al.} & 1979 & \cite{gottscho79a} & 1.9033125(240)  & 5.934(11) &-\\
        Miller \emph{et al.}  & 1984 & \cite{miller84b} & 1.903350(29) & 5.909(48) &- \\
        Rosner \emph{et al.}  & 1985 & \cite{rosner85a} & - &  - & 279.1(6)\\
        Berrah-Mansour \emph{et al.}  & 1991 & \cite{berrahmansour91a} &- & -& 276.92253(13) \\
        Ferguson \emph{et al.} & 1992  & \cite{ferguson92a} & 1.903410(10) & 5.970(8) & 277.1(9) \\
        Scholl \emph{et al.}  & 1992 & \cite{scholl92a} & 1.9033937(49) & 5.9813(40) & -\\
        Harada \emph{et al.}  & 1994 & \cite{harada94a} & 1.903403(70) & 5.979(17) &-\\
        Collet \emph{et al.}  & 1998 & \cite{collet98a} & 1.90330(2) & 5.904(21) & 274.5(1.3)\\
        Michaud \emph{et al.} & 2000 & \cite{michaud00a} & 1.9033800(45) & 5.9580(5) & 275.8(1.8)\\
        Widicus Weaver \emph{et al.} & 2008 & \cite{widicusweaver08a} & 1.903298(51) & 5.804(35) & 279.1(3.3)\\
        Zhang \emph{et al.} & 2015 & \cite{zhang15a} & 1.9034377(109) & 6.0031(168) & 277.67(1.02)\\
        \hline
        \multicolumn{3}{|c|}{Mean value } & 1.9033885(29) & 5.9583(5) &276.92253(13)\\
        \hline
    \end{tabular}
    \label{tab:mean_constants}
\end{table*}
\begin{table*}
    \centering
    \caption{\textbf{Fundamental vibrational frequencies $\Delta G_{10}$ of N$_2^+$.} Values reported in the literature compared to the present result. Values with asterisks were calculated from the equilibrium vibrational and anharmonicity constants quoted.}
    \begin{tabular}{|c|c|c||l|}
    \hline
        Author   & Year & Ref & $\Delta G_{10}$ [cm$^{-1}$] \\
        \hline
        \hline
        Gottscho \emph{et al.}* & 1979 & \cite{gottscho79a} & 2174.75(6)  \\
        Miller \emph{et al.}*  & 1984 & \cite{miller84b} & 2174.76(1)  \\
        Ferguson \emph{et al.}& 1992  & \cite{ferguson92a}  & 2174.7493(8)  \\
        Harada \emph{et al.}*   & 1994 & \cite{harada94a} & 2174.748(9)  \\
        Scholl \emph{et al.}*   & 1998 & \cite{scholl98a} & 2174.74(1) \\
        Al-Khalili \emph{et al.}*  & 1998 & \cite{alkhalili98a} & 2174.75(3)\\
        Michaud \emph{et al.} & 2000 & \cite{michaud00a} & 2174.746(1)\\
        Zhang \emph{et al.}   & 2015 & \cite{zhang15a} & 2174.743(1)\\
        This work & 2026 &  & 2174.74971(7)\\
        \hline
    \end{tabular}

    \label{tab:vib_constants}
\end{table*}

\begin{table*}
\centering
\caption{\textbf{Least-squares fit results.} Observed and fitted frequencies of the hyperfine components of the S(0) rovibrational band of N$_2^+$.}
\footnotesize
\begin{tabular}{|c||c|c|c|}
\hline
Transition & $(f_0-f')_{exp}$ [MHz] & $(f_0-f')_{fit}$ [MHz] & Residual [MHz] \\
\hline
\hline
$L1. \ket{F''=5/2} \rightarrow \ket{F'=5/2}$ & 876.5(14) &  876.4 &  0.1\\
$L2. \ket{F''=5/2} \rightarrow \ket{F'=7/2}$ & 932.8(14) & 932.9 & -0.1\\
$L3. \ket{F''=5/2} \rightarrow \ket{F'=9/2}$ & 983.4(20) & 983.1 & 0.3\\
$L5. \ket{F''=3/2} \rightarrow \ket{F'=1/2}$ & 1038.5(20) & 1038.4 & 0.1\\
$L6. \ket{F''=3/2} \rightarrow \ket{F'=3/2}$ & 1077.6(14) & 1077.6 & 0.0\\
\hline
\end{tabular}
\label{tab:hf_fit}
\end{table*}

\begin{table*}
\caption{\textbf{Values of the hyperfine constants} obtained from the least-squares fit of the hyperfine components of the S(0) rovibrational band of N$_2^+$ and previously reported values.}
    \centering
    \begin{tabular}{|c||c|c||c|c|c|}
    \hline
        Constant & Fit & Literature & Reference\\
        \hline\hline
        $b_{F0}$ [MHz] & 101.8(10) & 102.4(11) &~\cite{scholl98a}\\
        $b_{F1}$ [MHz] & 102.5(18) & 100.6040(15) &~\cite{berrahmansour91a}\\
        $t_{1}$ [MHz] & 29.9(14) & 28.1946(13) &~\cite{berrahmansour91a}\\
        \hline
    \end{tabular}
    \label{tab:hf_constants_fitted}
\end{table*}

\clearpage
\bibliography{Bibliography}

\end{document}